\documentclass[12pt]{article}
\usepackage{amsmath}
\usepackage{graphicx}
\usepackage{enumerate}
\usepackage{natbib}
\usepackage{url} 

\newcommand{\blind}{1}

\usepackage[utf8]{inputenc}
\usepackage[english]{babel}
\usepackage{amsmath}
\usepackage{amsfonts}
\usepackage{amssymb, color}
\usepackage{graphicx}
\usepackage{fullpage}
\usepackage{natbib}
\usepackage{amsthm}
\usepackage{dsfont}
\usepackage{multirow}

\newcounter{lm}

\newtheorem{lemma}[lm]{Lemma}

\DeclareMathOperator*{\argmax}{arg\,max}
\DeclareMathOperator*{\argmin}{arg\,min}

\newcommand{\tit}{Simultaneous semi-parametric estimation of clustering and regression}

\newcommand{\rU}{\mathbb{U}}

\newcommand{\rX}{\mathbb{X}}
\newcommand{\rZ}{\mathbb{Z}}
\newcommand{\rY}{\mathbb{Y}}

\usepackage{tikz}
\usetikzlibrary{bayesnet}
\usepackage{authblk}
\begin{document}

\def\spacingset#1{\renewcommand{\baselinestretch}%
{#1}\small\normalsize} \spacingset{1}


%

\if1\blind
{
  \title{\bf \tit}
\author[1]{Matthieu Marbac}
\author[2]{Mohammed Sedki}
\author[3]{Christophe Biernacki}
\author[4]{Vincent Vandewalle}
\affil[1]{Univ. Rennes, Ensai, CNRS, CREST   - UMR 9194, F-35000 Rennes, France}
\affil[2]{Univ. Paris-Sud and Inserm, France}
\affil[3]{Inria, Univ. Lille, CNRS, UMR 8524 - Laboratoire Paul Painlevé, F-59000 Lille}
\affil[4]{Univ. Lille, CHU Lille, ULR 2694 - METRICS : Évaluation des technologies de santé et des pratiques médicales, F-59000 Lille, France}
  \maketitle
} \fi

\if0\blind
{
  \bigskip
  \bigskip
  \bigskip
  \begin{center}
    {\LARGE\bf \tit}
\end{center}
  \medskip
} \fi

\bigskip
\begin{abstract}
We investigate the parameter estimation of regression models with fixed group effects, when the group variable is missing while group related variables are available. This problem involves clustering to infer the missing group variable based on the group related variables, and regression to build a model on the target variable given the group and eventually additional variables. Thus, this problem can be formulated as the joint distribution modeling of the target and of the group related variables. The usual parameter estimation strategy for this joint model is a two-step approach starting by learning the group variable (clustering step) and then plugging in its estimator for fitting the regression model (regression step). However, this approach is suboptimal (providing in particular biased regression estimates) since it does not make use of the target variable for clustering. Thus, we claim for a simultaneous estimation approach of both clustering and regression, in a semi-parametric framework.
Numerical experiments illustrate the benefits of our proposition by considering wide ranges of distributions and regression models. The relevance of our new method is illustrated on real data dealing with problems associated with high blood pressure prevention.
\end{abstract}
\noindent%
\vfill

\newpage
\spacingset{1.5} 


\section{Introduction}
Regression models allow the relationship between some covariates and a target variable to be investigated. 
These models are defined by an equation on the conditional moment of a transformation of the noise. 
This transformation is generally the piecewise derivative of the loss function that defines the type of regression: mean, robust,  quantile \citep{koenker1978regression,horowitz2005nonparametric,wei2009quantile}, expectile  \citep{newey1987asymmetric,ehm2016quantiles,daouia2018estimation}.

\medskip

The regression model with a fixed group effect is central within this generic paradigm. It considers that the intercept of the regression depends on the group from which the subject belongs (the intercept is common for subjects belonging to the same group but different for subjects belonging to different groups). However, in many applications, the group variable is not observed but other variables related to this variable are observed. 
For instance, suppose we want to investigate high blood pressure by considering the levels of physical activity among the covariates.  In many cohorts, the level of physical activity of a subject is generally not directly available (because such a  variable is not easily measurable) but many variables on the mean time spent doing different activities are available. 
Note that the regression model with a fixed group effect and a latent group variable is a specific mixture of regressions \citep{wang1996mixed,hunter2012semiparametric,wu2016mixtures} where only the intercepts of the regressions are different among the components and where the mixture weights depend on some other variables. 
Moreover,  the regression model with a fixed group effect and a latent group variable can be interpreted as a regression model with specific quantization of the variables that we use to estimate the group membership (see for instance \citet{CHARLIER201514} for the quantization in quantile regression).

\medskip

The estimation of a regression model with a fixed group effect is generally performed using a \emph{two-step approach} as for instance in epidemiology or in economics \citep{auray2015clustering,ando2016panel,zhang2019quantile}. As a first step, a clustering on the individual based on the group related variables is performed to obtain an estimator of the group. As a second step, the regression model is fit by using the estimator of the group variable among the covariates. 
The second step considers a regression model with measurement errors on the covariates. Indeed, the group variable is estimated in the clustering step with errors. Hence, it is well-known that the resulting estimators of the parameters of regression are biased (see for instance \citet{carroll1991semiparametric, nakamura1992proportional, bertrand2017robustness}). The bias depends on the accuracy of the clustering step. Note that, despite that the target variable contains information about the group variable (and so is relevant for clustering), this information is not used in the two-step approach, leading to a suboptimal procedures.

\medskip

Some simultaneous approaches have been considered in the framework of latent variable models, such as latent class and latent profile analysis \citep{guo2006latent,kim2016modeling}. In this framework, the authors introduce latent class and latent factor variables to explain the heterogeneity of observed variables. However no direct focus is taken to explain a particular variable given other ones, and the approach is limited to a parametric framework. Another related reference is the work of \cite{sammel1997latent}, where the authors introduce a latent variable mixed effects model, which allows for arbitrary covariate effects, as well as direct modelling of covariates on the latent variable. Some other relevant references can be found in the field of concomitant variables~\citep{dayton_concomitant-variable_1988,grun_flexmix_2008,vankatova_evaluation_2017}, where some additional variables are used to locally adjust the weights of the mixture of regressions. But, these approaches are rather focused on the mixture of regressions task than on clustering data based on concomitant variables. 

\medskip

We propose a new procedure (hereafter referred to as the \emph{simultaneous approach}) that estimates simultaneously the clustering and the regression models in a semi-parametric frameworks  \citep{hunter2011nonparametric} thus circumventing the limits of the standard procedure (biased estimators). We demonstrate that this procedure improves both the estimators of the partition and regression parameters. A full parametric setting is also presented, however if one of the clustering or regression model were ill-specified, its bias modeling could contaminate the results of the other one.
Thus we focus on semi-parametric mixture where the component densities are defined as a product of univariate densities \citep{chauveau2015semi,zhu2016theoretical,zhengJASA2019}, which is identifiable if the univariate densities are linearly independent and if at least  three variables are used for clustering \citep{allman2009identifiability}. Note that, mixtures of symmetric distributions \citep{hunterAOS2007,butuceaScand2014} could also be considered in a similar way. 
Semi-parametric inference is achieved by a maximum smoothed likelihood approach \citep{levine2011maximum} via a Maximization-Minimization (MM) algorithm \citep{hunter2004tutorial}. Note that selecting the number of components in a semi-parametric mixture is not easy \citep{kasahara2014non,kwon2019estimation}. However, in our context, the number of components can be selected according to the quality of the prediction of the target variable.

\medskip

This paper is organized as follows.
Section~\ref{sec:grl} introduces a general context where a statistical analysis requires both methods of clustering and prediction, and it presents the standard approach that estimates the parameters in two steps. 
Section~\ref{sec:unified} shows that a procedure that allows a simultaneous estimation of the clustering and of the regression parameters generally outperforms the two-step approach. This section also briefly presents the simultaneous procedure on parametric framework, then focuses on the semi-parametric frameworks. 
Section~\ref{sec:simu} presents numerical experiments on simulated data  showing the benefits of the proposed approach. 
Section~\ref{sec:ill} illustrates our proposition for  problems associated with high blood pressure prevention. 
Section~\ref{sec:conclusion} provides a conclusion and discussion about extensions. The mathematical details are presented in Appendix~A.

\section{Embedding clustering and prediction models} \label{sec:grl}

\subsection{Data presentation}
Let $(V^\top,X^\top,Y)^\top$ be the set of the random variables where $V=(U^\top,Z^\top)^\top$ is a $d_V=d_U + K$ dimensional vector used as covariates for the prediction of the univariate variable $Y\in\mathbb{R}$, $X$ is a $d_X$ dimensional vector and $Z=(Z_1,\ldots,Z_K)^\top\in\mathcal{Z}$ is a categorical variable with $K$ levels.  The variable $Z$ indicates the group membership such that $Z_k=1$ if the subject belongs to cluster $k$  and otherwise $Z_k=0$. The realizations of $(U^\top,X^\top,Y)^\top$ are observed but the realizations of $Z$ are unobserved. Thus, $X$ is a set of proxy  variables used to estimate the realizations of $Z$. Considering the high blood pressure example, $Y$ corresponds to the  diastolic blood pressure, $U$ is the set of  observed covariates (gender, age, alcohol consumption, obesity and  sleep quality), $X$ is the set of  covariates measuring the level of physical activity and $Z$ indicates the membership of a group of subjects with similar physical activity behaviours. The observed data  are $n$ independent copies of $(U^\top,X^\top,Y)^\top$ denoted by  $\rU=(u_1,\ldots,u_n)^\top$,  $\rX=(x_1,\ldots,x_n)^\top$ and  $\rY=(y_1,\ldots,y_n)^\top$ respectively.  The $n$ unobserved realizations of $Z$ are denoted by $\rZ=(z_1,\ldots,z_n)^\top$.

\subsection{Motivating example}
We use the following example throughout the paper, which examines the general objective of high blood pressure prevention. Here, we focus on the detection of indicators related to the diastolic blood pressure ($Y$); see \citet{berney2018isolated} for the interest of the study. The indicators we wish to consider are the gender, the age, the alcohol consumption, the obesity, the sleep quality and the level of physical activity ($V$). However, the level of physical activity ($Z$) of a patient is not directly measured and we only have a set of variables which describes the physical activity ($X$), such as practice of that recreational activity, hours spent watching TV, hours spent on the computer, {\it etc.} More details of the data are provided in Section~\ref{sec:ill}. The study of the different indicators is performed using a regression model that explains the diastolic blood pressure with a set of covariates where one variable (the physical activity) was not directly observed. Information about this latter variable is available from other variables that do not appear in the regression.

\subsection{Introducing the joint predictive clustering model} \label{sec:model}

\paragraph{Regression model}
Let a loss function be  $\mathcal{L}(\cdot)$ and $\rho(\cdot)$ its piecewise derivative. The loss function $\mathcal{L}$ allows  the regression model of $Y$  on $V$ to  be specified with a fixed group effect given by
\begin{equation} \label{eq:modelpred}
Y = V^\top \beta + \varepsilon \text{ with }  \mathbb{E}[\rho(\varepsilon)| V]=0,
\end{equation}
where $\beta=(\gamma^\top,\delta^\top)^\top \in\mathbb{R}^{d_V}$, $\gamma\in\mathbb{R}^{d_U}$ are the coefficients of $U$, $\delta=(\delta_1,\ldots,\delta_K)^\top\in \mathbb{R}^K$ are the coefficients of $Z$ (\emph{i.e.,} the parameters of the group effect), and $\varepsilon$ is the noise. Note that for reasons of identifiability, the model does not have an intercept. The choice of $\mathcal{L}$ allows many models  to be considered and, among them, one can cite the mean regression (with $\mathcal{L}(t)=t^2$ and $\rho(t)=2t$), the $\tau$-quantile regression (with $\mathcal{L}(t)= |t|+(2\tau-1)t$ and $\rho(\varepsilon)=\tau - \mathds{1}_{\{\varepsilon \leq 0\}} $; \citet{koenker1978regression}), the $\tau$-expectile regression (with $\mathcal{L}(t)=|\tau - \mathbf{1}\{t\leq 0\}|t^2$ and  $\rho(t) = 2t( (1-\tau) \mathbf{1}\{t\leq 0\} + \tau \mathbf{1}\{t > 0\} )$; \citet{newey1987asymmetric}), {\it etc.}

The restriction on the conditional moment of $\rho(\varepsilon)$ given $V$ is sufficient to define a model and allows for parameter estimation. However, obtaining  maximum likelihood estimate (MLE) needs specific assumptions on the noise distribution.  For instance, parameters of the mean regression can be consistently estimated with MLE by assuming a centred Gaussian noise. Similarly, the parameters of  $\tau$-quantile (or $\tau$-expectile) regression  can be consistently estimated with MLE by assuming that the noise follows an asymmetric Laplace  (or an asymmetric normal) distribution \citep{yu2001bayesian,xing2017bayesian}. Hereafter, we denote the density of the noise $\varepsilon$  by $f_\varepsilon$.

\paragraph{Clustering model}
The distribution of $X$ given $Z_k=1$ is defined by the density  $f_k(\cdot)$. Therefore, the marginal distribution of $X$ is a mixture model defined by the density
\begin{equation} \label{eq:modelclust}
f(x;\vartheta) = \sum_{k=1}^K \pi_k f_k(x),
\end{equation} 
where $\vartheta=\{\pi_k,f_k;k=1,\ldots,K\}$, $\pi_k>0$ and $\sum_{k=1}^K \pi_k=1$ and where $f_k$ is the density of component $k$. In a parametric approach, $f_k$ is assumed to be parametric so it is denoted by $f_k(\cdot;\alpha_k)$ where $\alpha_k$ are the parameters of component $k$. In a semi-parametric approach, some assumptions are required to ensure  model identifiability (see for instance \citet{chauveau2015semi}). In the following, the semi-parametric approaches are considered with the assumption that each $f_k$ is a product of univariate densities (see Section~\ref{sec:np}). 

\paragraph{Joint clustering and regression model}
The joint model assumes that $Z$ explains the dependency between $Y$ and $X$ (\emph{i.e.,} $Y$ and $X$ are conditionally independent given $Z$) and that  $U$ and $(X^\top,Z^\top)$ are independent. 
Moreover, the distribution of $(X,Y)$ given $U$ is also a mixture model defined by the density (noting $\theta=\{\vartheta\}\cup\{\delta_k;k=1,\ldots,K\}\cup\{\gamma,f_\varepsilon\}$)
\begin{equation} \label{eq:modeljoint}
f(x,y|u;\theta) = \sum_{k=1}^K \pi_k f_k(x) f_\varepsilon(y - u^\top\gamma - \delta_k),
\end{equation}
where, for $k=1,\ldots,K$ we have 
\begin{equation}
\mathbb{E}[\rho(Y - U^\top\gamma - \delta_k)| U,Z_k=1]=0,\label{eq:condmoment}
\end{equation}

Note that \eqref{eq:modeljoint} is a particular mixture of regressions model where the mixture weights are proportional to $\pi_k f_k(x)$ (thus depending on covariates that do not appear in the regressions) and where only the intercepts (\emph{i.e.,} $\delta_1,\ldots,\delta_K$) are different among the regressions. Contrary to~\citet{grun_flexmix_2008} who consider the density $f(y|u,x;\theta)$ thus focusing on the regression framework, here we propose to consider the density $f(y,x|u;\theta)$ which balances the regression and the clustering frameworks.

%
%
%

\paragraph{Moment condition}
The following lemma gives the moment equation verified on the joint model. It will be used later to justify the need for a simultaneous approach. 


\begin{lemma} \label{lemma:eqmoment}
Let an identifiable model defined by \eqref{eq:modeljoint} and \eqref{eq:condmoment}, where, for any $x$ and $k$. Then, noting  
$r^{X,Y}_k(x,y)=\frac{\pi_k f_k(x) f_\varepsilon(y - u^\top\gamma - \delta_k)}{\sum_{\ell=1}^K \pi_\ell f_\ell(x) f_\varepsilon(y - u^\top\gamma - \delta_\ell)}$, $\beta=(\delta^\top,\gamma^\top)^\top$
is the single parameter satisfying
\begin{equation}
\forall k=1,\ldots,K,\; \mathbb{E}[r^{X,Y}_k(X,Y)\rho(Y - u^\top\gamma - \delta_k)|U,X]=0. \label{eq:eqmoment}
\end{equation}
\end{lemma}

\section{The proposed simultaneous estimation procedure} \label{sec:unified}
  
\subsection{Limits of the standard two-step approach estimation} \label{subsec:twostep}
The aim is to explain the distribution of $Y$ given $V=(U^\top,Z^\top)^\top$ from an observed sample. A direct estimation of the model \eqref{eq:modelpred} is not doable because the realizations of $Z$ are unobserved. The standard approach considers the following  two-steps:
\begin{enumerate}
\item {\textbf{Clustering step}} Perform a clustering of $\rX$ to obtain an estimated hard classification rule $\hat r^X:\mathbb{R}^{d_{X}} \rightarrow \mathcal{Z}$  or an estimated fuzzy  classification rule $\hat r^X:\mathbb{R}^{d_{X}} \rightarrow \tilde{\mathcal{Z}}_K$ where $\tilde{\mathcal{Z}}_K$ is the simplex of size $K$.  
\item {\textbf{Regression step}} Estimation of the regression parameters given the estimator of the group memberships $\hat\beta^{\hat r^X}:=(\hat\gamma^{\hat r^X \top}, \hat\delta^{\hat r^X \top})^\top = \argmin_{\beta} \sum_{i=1}^n\sum_{k=1}^K \hat r^X_k(x_i)\mathcal{L}(y_i - u_i^\top \gamma - \delta_k)$
where $\hat r^X_k(x_i)$ is the element $k$ of vector $\hat r^X(x_i)$. Note that $\hat r^X_k(x_i)$ is an estimator of the conditional probability that observation $i$ belongs to cluster $k$ given $x_i$, if the fuzzy classification rule is used. 
\end{enumerate}

 The following lemma states that the two-step approach produces asymptotically biased estimators of the partition and regression parameters, even if the optimal classification rule on $X$ is used. 
 
\begin{lemma}\label{lemm:resasymptotic}
Let an identifiable model defined by \eqref{eq:modeljoint} and \eqref{eq:condmoment}, then
\begin{enumerate}
\item If classification rule $\hat r^X$ converges to a classification rule $\tilde r^X$ (\emph{i.e.,} the best classification rule based on $X$), thus $\hat r^X$ is asymptotically suboptimal since
$\mathbb{E}\left[\sum_{k=1}^K \tilde r^X_k(X)Z_k\right]<\mathbb{E}\left[\sum_{k=1}^K r^{X,Y}_k(X,Y)Z_k\right]$, 
with $\tilde{r}_k^X(x)\propto\pi_k f_k(x)$.   
\item  Let $f_\varepsilon$ defines a random variable with finite variance and $U$ to have a covariance matrix with non-zero eigenvalues. Then, considering the quadratic loss and a consistent fuzzy classification rule $\hat r^X$, the estimator $\hat\gamma^{\hat r^X}$ is asymptotically unbiased but the estimator $\hat\delta^{\hat r^X}$ is asymptotically biased since
$\lim_{n\to\infty}\text{bias}( \hat\delta_k^{\hat r^X}) =\frac{\sum_{\ell=1}^K \Delta_{k\ell} \delta_\ell}{\sum_{h=1}^K \Delta_{kh}} - \delta_k$,
where $\Delta_{k\ell}=\mathbb{E}[r_k^X(X)r_\ell^X(X)]$.
\end{enumerate}
\end{lemma}

Thus the clustering step provides a suboptimal classification rule because the classification neglects the information given by $Y$. Consequently, the regression step provides estimators that are asymptotically biased and implies fitting the parameters of a regression model with measurement errors in the covariates (for instance, considering the hard assignment, we have no guarantee of having a perfect recovery of the partition, \emph{i.e.,},  $\hat r^X(x_i) = z_i$, for $i=1,\ldots,n$). The measurement errors generally produce biases in the estimation. Finally, the quality of the estimated classification rule directly influences the quality of the estimator of the regression parameters.

\subsection{Limits of a parametric simultaneous procedure} \label{sec:param}
In this section, we consider a probabilistic approach with a parametric point-of-view. Thus, the family of distributions of each component $k$ is supposed to be known and parameterized by $\alpha_k$. Moreover, the distribution of the noise $f_\varepsilon$ is chosen according to the type of the regression under consideration (see the discussion in Section~\ref{sec:model}).
The aim of the simultaneous procedure can be achieved by maximizing the log-likelihood of $\rY,\rX$ given $\rU$ with respect to~$\theta$
$$
\ell(\theta;\rY,\rX\mid \rU) = \sum_{i=1}^n \ln\left(
\sum_{k=1}^K \pi_k f_k(x_i;\alpha_k) f_\varepsilon(y_i -  u_i^\top \gamma - \delta_k)
\right),
$$ 
where $\theta=(\pi_1,\ldots,\pi_K,\alpha_1^\top,\ldots,\alpha_K^\top,\beta^\top)$ groups all the model parameters. Indeed, the maximum likelihood inference using $\ell(\theta;\rY,\rX\mid \rU) $ allows for similarly learning the classification rule based on $(X,Y)$ and the regression coefficients. This function cannot be directly maximized so we consider the complete-data log-likelihood with data $(\rY,\rX,\rZ)$ given $\rU$ defined by
$$
\ell(\theta;\rY,\rX, \rZ\mid \rU) = \sum_{i=1}^n \sum_{k=1}^K z_{ik} \ln\left(
\pi_k f_k(x_i;\alpha_k) f_\varepsilon(y_i -  u_i^\top \gamma - \delta_k)
\right).
$$
The MLE $\hat\theta$ can be obtained via an EM algorithm presented in Appendix~\ref{sec:EM}. 
Let an identifiable model defined by \eqref{eq:modeljoint} and \eqref{eq:condmoment}, then
\begin{enumerate}
\item If all the parametric distributions are well-specified, then properties of the MLE imply that the classification rule is asymptotically optimal and $\hat\beta$ is asymptotically unbiased.
\item If at least one parametric distribution is misspecified, then the classification rule is generaly asymptotically suboptimal and  $\hat\beta$ is generally asymptotically biased.
\end{enumerate}
  
Let notice that the distribution of the noise appears at the E-step and thus influences the classification rule. Hence, the classification rule is deteriorated if the distribution of the noise is misspecified. This is not the case when estimation is performed using the two-step approach, since clustering is performed prior to regression, and regression can still be unbiased if the moment condition (see Lemma~\ref{lemma:eqmoment}) is well-specified. Thus, in the next section, we propose a semi-parametric approach that circumvents this issue because it does not assume a specific family of distributions for the noise and the components.

\subsection{Advised simultaneous semi-parametric procedure} \label{sec:np}

\paragraph{Semi-parametric model}
In this section, we consider the semi-parametric version of model defined by \eqref{eq:modeljoint} where the densities of the components are assumed to be a product of univariate densities. Thus, we have
\begin{equation}\label{eq:npmodel}
f(y,x \mid u ; \theta ) = \sum_{k=1}^K \pi_k f_k(y,x \mid u ; \theta ) =  \sum_{k=1}^K \pi_k \prod_{j=1}^{d_X} f_{kj}(x_j) f_{\varepsilon} (y - u^\top \gamma - \delta_k),
\end{equation}
where $\theta$ groups all the finite and infinite parameters and $\beta$ is such that \eqref{eq:eqmoment} holds.
 A sufficient condition implying identifiability for model  \eqref{eq:npmodel} is that the marginal distribution of $X$ is identifiable and thus a sufficient condition is to consider linearly independent densities $f_kj$'s and $d_X\geq 3$ \citep{allman2009identifiability}.  For sake of simplicity we will note $w = (x^\top,y)^{\top}$  with $w \in \mathbb{R}^{d_X + 1}$,  such that $f(y,x \mid u ; \theta )=  \sum_{k=1}^K \pi_k f_k(w \mid u ; \theta ) $.

\paragraph{Smoothed log-likelihood}
Let $\mathcal S$  be the smoothing operator defined by
$\mathcal S f_k(w\mid u;\theta) = \int K_h(w - \tilde w) f_k(\tilde w\mid u; \theta) d\tilde w$, where $K_h(a)=\prod_{j=1}^d K_h(a_j)$ with $a\in\mathbb{R}^d$ and with $K_h(a_j)$ is a rescale kernel function defined by $K_h(a_j) = h^{-1} K(h^{-1}a_j)$ where $h$ is the bandwidth. 
For the semi-parametric approach, the estimation can be performed by maximizing the smoothed log-likelihood \citep{levine2011maximum} defined by $\ell(\theta) = \sum_{i=1}^n \ln \left( \sum_{k=1}^K \pi_k \left(\mathcal N f_k \right)(w\mid u; \theta) \right)$
subject to the empirical counterpart of \eqref{eq:eqmoment}, namely
$\frac{1}{n} \sum_{i=1}^n \sum_{k=1}^K \frac{f_k(w\mid u;\theta)}{\sum_{\ell=1}^K \pi_\ell f_\ell(w\mid u;\theta)} \rho(y_i - u_i^\top\gamma - \delta_k) = 0$,
where $\left(\mathcal N f_k \right)(w\mid u; \theta) = \exp \left\{\mathcal S \ln f_k(w\mid u;\theta) \right\}=\exp \left\{ \int K_h(w - \tilde w) \ln f_k(\tilde w\mid u; \theta) d\tilde w) \right\} $.

\paragraph{Majorization-Minorization algorithm} Parameter estimation is achieved via a Majorization-Minorization algorithm. Given an initial value $\theta^{[0]}$, this algorithm iterates between a majorization and a minorization step. Thus, an iteration $[r]$ is defined by
\begin{itemize}
\item Majorization step:
$ t_{ik}^{[r-1]} \propto \pi_k^{[r-1]} \left(\mathcal N f_k^{[r-1]} \right)(w_i\mid u_i; \theta^{[r-1]}) .
$
\item Minorization step:
\begin{enumerate}
\item Updating the parametric elements
$$\pi^{[r]}_k = \frac{1}{n} \sum_{i}  t_{ik}^{[r-1]} \text{ and }
\beta^{[r]} = \argmin_{\beta} \sum_{i,k} t_{ik}^{[r-1]} \rho(y_i - u_i^\top \gamma - \delta_k).$$
\item Updating the nonparametric elements
$$
f_{kj} (a) = \frac{1}{n \pi_k^{[r]}} \sum_{i} t_{ik}^{[r-1]} K_h(x_{ij} - a)
\text{ and }
f_{\varepsilon} (a) = \frac{1}{n} \sum_{i,k} t_{ik}^{[r-1]} K_h(y_i - u_i^\top \gamma^{[r]} - \delta_k^{[r]} - a).
$$
\end{enumerate}
\end{itemize}

The Majorization-Minorization algorithm is monotonic for the smoothed log-likelihood. It is a direct consequence of the monotony of the algorithm of \citet{levine2011maximum} where we use the fact that, in order to satisfy the moment condition defined in \eqref{eq:eqmoment} of Lemma~\ref{lemma:eqmoment}, we must have $\beta^{[r]} = \argmin_{\beta} \sum_{i=1}^n \sum_{k=1}^K t_{ik}^{[r-1]} \rho(y_i - u_i^\top \gamma - \delta_k)$.

As in  \citet{hunter2012semiparametric}, the majorization step is not explicit. However, because it only implies univariate integrals, it can be efficiently assessed by numerical approximations. Finally, bandwidth selection can be performed as usual for semi-parametric mixtures (see \citet{chauveau2015semi}). However, like in any supervised problem, we can use the cross-validated accuracy of the prediction of $Y$ for bandwidth selection. 
 
\section{Numerical experiments}\label{sec:simu}

\subsection{Simulation setup} \label{sec:simusetup}
Data are generated from a bicomponent mixture with equal proportions (\emph{i.e.,} $K=2$ and $\pi_1=\pi_2=1/2)$ such that the density of $X_i$ given $Z_i$, is a product of univariate densities. For $j=1,\ldots,4$, $X_{ij}= (-1)^{Z_{i1}} \xi + \eta_{ij}$ where the group variable $Z_i$ is sampled according to a multinomial distribution of parameters $\pi_1$ and $\pi_2$, and $\eta_{ij}$  are independent and identically distributed random variables. 
Moreover, we have $U_i\sim\mathcal{N}_2(0,\mathbf{I}_2)$ and $Y_i~=~(Z_{i1},Z_{i2},U_{i1},U_{i2})\beta + \varepsilon_i,$ with $\beta = (-1,1,1,1)^\top$. Different distributions are used for $\eta_{ij}$ and $\varepsilon_i$. $100$ samples are generated for each case, and the parameter $\xi$, controlling the class overlap,  is tuned to have a theoretical misclassification of 0.10. The approaches are compared with the MSE of the coefficient of the regression $\beta$ and the adjusted rand index (ARI) between the true and the estimated partition. The semi-parametric approach is applied with a fixed bandwidth $h=n^{-1/5}$, this choice is considered for sake of simplicity, the tuning of this window could be considered as in~\cite{chauveau2015semi}.

\subsection{Method comparison} \label{sec:simucompare}
In this experiment, we show that the simultaneous procedure outperforms the standard two-step procedure, in both parameteric and semi-parametric framework. In the case where the parametric model is well-specified, we show that its results are equivalent to those obtained by the semi-parametric model. Moreover, we show that if one distribution is misspecified (component distribution or noise distribution), the results of the parametric approach are deteriorated even if the moment condition of the regression model is well-specified. Thus, we advise to use the semi-parametric model if the family of the distributions is unknown to prevent the bias in the estimation.

\begin{figure}[ht!]
\centering \includegraphics[scale=0.45]{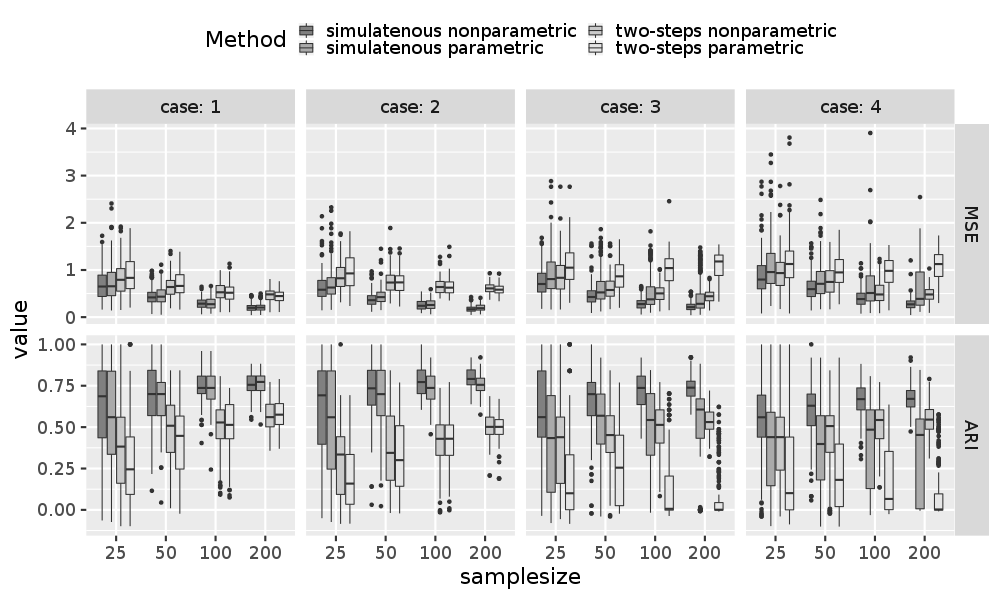} 
\caption{\label{fig:compare}Boxplots of the MSE of the estimators of the regression parameters and ARI obtained by the simultaneous and two-step methods, in a parametric and semi-parametric framework, on 100 samples of size $n$.}
\end{figure}

In this experiment, we focus on the quadratic loss. We consider one case where the parametric model is well-specified (case-1) where $\eta_{ij}$ and $\varepsilon_i$ follow standard Gaussian distributions. We also consider one case where the parametric model is misspecified on the noise distribution (case-2) where $\eta_{ij}$ follows a standard Gaussian distribution and $\varepsilon_i=\tau_i -1$ with $\tau_i\sim\mathcal{E}xp(1)$, and one case where the parametric model is misspecified on the component distributions (case-3) where $\eta_{ij}$ follows a  Student distribution with 3 degrees of freedom and $\varepsilon_i$ follows a standard Gaussian distribution. Finally, we consider one case where the parametric model is misspecified on both of the component and noise distributions (case-4), where $\eta_{ij}$ and $\varepsilon_i$ follows Student distributions with 3 degrees of freedom. Figure~\ref{fig:compare}  shows the MSE of the regression parameters and the ARI obtained.

\subsection{Robust regression} \label{sec:simurobust} 
When the noise of a regression follows an heavy-tail distribution, robust regressions (median, Huber and logcosh regressions) allow for improvement of the estimators of the regression coefficients compare to the ordinary least square estimators. Despite that, with a suitable assumption on the noise distribution, the simultaneous parametric approach could consider such regressions, the parameteric assumptions made on the noise distribution would be quite unrealistic (\emph{e.g.,} Laplace distribution for the median regression). Thus, we now illustrate that the simultaneous approach can easily consider robust regressions, in a semi-parametric framework, and that the resulting estimators are better than those obtained with the quadratic loss. 

\begin{figure}[ht!]
\centering \includegraphics[scale=0.45]{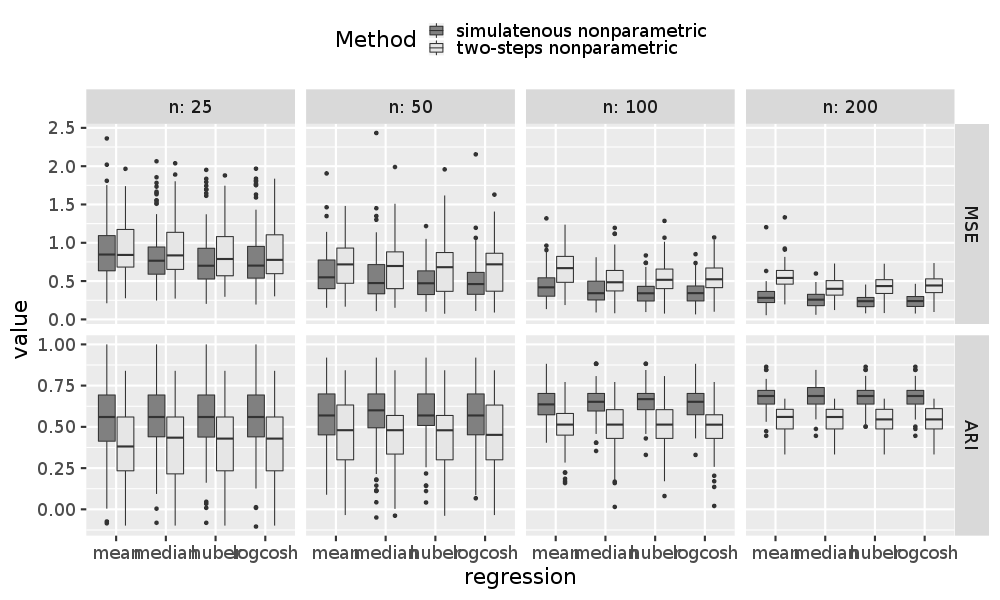} 
\caption{\label{fig:robust} Boxplots of the MSE of the estimators of the regression parameters and ARI obtained by different regressions (mean, median, Huber with parameter 1 and logcosh) estimated with the simultaneous and two-step semi-parametric methods on 100 samples of size $n$ generated with heavy-tail distributions.}
\end{figure}

In this experiment, we consider that $\eta_{ij}$ and $\varepsilon_i$ follow independent Student distributions with three degrees of freedom. Figure~\ref{fig:robust} shows the MSE of the regression parameters and the ARI obtained by the mean regression, the median regression, the Huber regression with parameter 1 and the logcosh regression. Results show that the simultaneous approach improves the estimators (according to the MSE and the ARI) for any type of regression and any sample size. Moreover, robust regressions improve the accuracy of the estimator of the regression parameters. However, for this simulation setup, this improvement does not affect the accuracy of the estimated partitions.

\subsection{Asymetric losses} \label{sec:simuasym}
\begin{figure}[ht!]
\centering \includegraphics[scale=0.45]{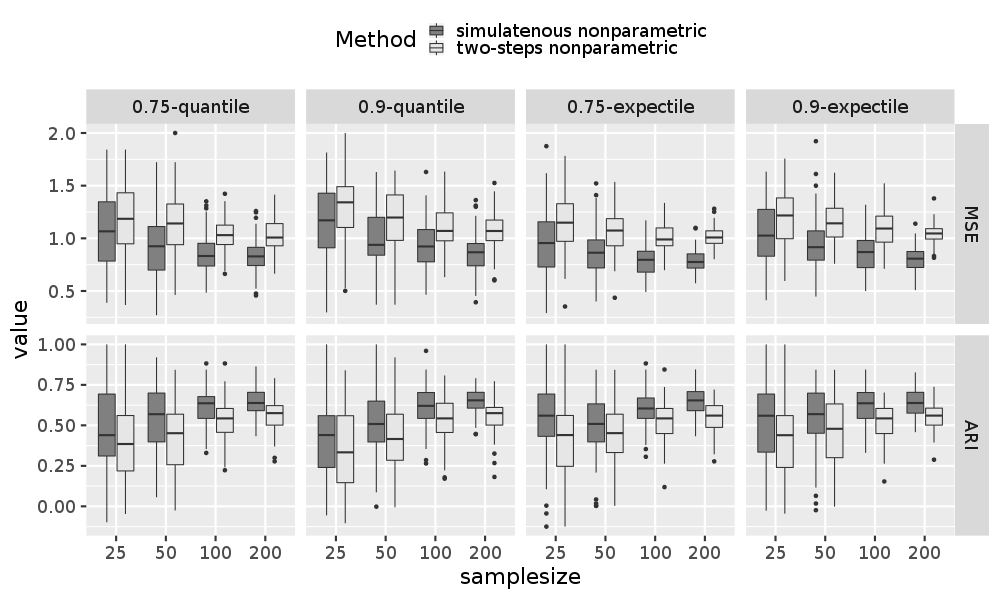} 
\caption{\label{fig:asym}Boxplots of the MSE of the estimators of the regression parameters and ARI obtained by asymetric regressions estimated with the simultaneous and two-step semi-parametric methods on 100 samples of size $n$.}
\end{figure}

Expectile and quantile regressions respectively generalize the mean and the median regression by focusing on the tails of the distribution of the target variable given the covariates. To illustrate the fact that the semi-parametric simultaneous method allows for easily considering these regression models, data are generated such that $\eta_{ij}\sim\mathcal{N}(0,1)$ and $\varepsilon_i\sim\mathcal{N}(-c_\tau,1)$. The scalar $c_\tau$ is defined according to the regression model. Thus, $c_\tau$ is the 0.75-expectile, 0.9-expectile, 0.75-quantile and 0.9-quantile for the 0.75-expectile, 0.9-expectile, 0.75-quantile and 0.9-quantile regression respectively. Figure~\ref{fig:asym} shows that the simultaneous semi-parametric approach improves the estimators compare to those provided by the two-step approach.

\section{High blood pressure prevention data set} \label{sec:ill}
\paragraph{Problem reminder}
We come back now to the problem of high blood pressure prevention. We focus on the detection of indicators related to the diastolic blood pressure. The indicators we want to consider are the gender, age, alcohol consumption,  obesity,  sleep quality and level of physical activity. However, the level of physical activity of a patient is not directly measured and we only have a set of variables that describe the physical activity. Thus, we want to cluster the subjects based on this set of variables to obtain patterns of similar physical activities and we want to use these patterns in the prediction of the diastolic blood pressure.

\paragraph{Material and methods}
The data were obtained from National Health and Nutrition Examination Survey of 2011-2012\footnote{The data are freely downloadable at \\  \emph{https://wwwn.cdc.gov/nchs/nhanes/continuousnhanes/default.aspx?BeginYear=2011}}. The target variable is the \emph{diastolic blood pressure} in mmHg (code BPXDI1). The seven covariates in $U$ are \emph{gender} which was equal to 1 for men et 0 for women (code RIAGENDR), \emph{age} (RIDAGEYR), \emph{alcohol} which indicates whether the subjects consume more than five drinks (for men) and four drinks (for women) of alcoholic beverages almost daily (computed from code ALQ151 and ALQ155), \emph{obesity} which indicates if the body mass index is more than 30 (computed from code BMXBMI), \emph{sleep} which indicates the number of hours of sleeping (computed from code SLD010H), \emph{smoke} which indicates if the subjects used tobacco/nicotine in the last five days (code SMQ680) and \emph{cholesterol} which indicates the total cholesterol in mg/dL (code LBXTC). All the subjects that had missing values for those variables were removed. Seven variables are used in $X$ to evaluate the level of physical activity. Among these variables, five variables are binary and indicated whether the subject has a vigorous work activity (code PAQ605),   whether the subject has a moderate work activity (code PAQ620),  whether the subject usually travels on foot or by bike (code PAQ635),  whether the subject has vigorous recreational activities (code PAQ650) and whether the subject has moderate recreational activities (code PAQ665). The two remaining variables in $X$ have 7 levels and indicate   the time spent   watching TV (code PAQ710) and the time spent using a computer (code PAQ715).  Finally, the studied population is composed of 2626 subjects between 18 and 60 years old. To investigate the performances of the different models, 67\% of the sample  (\emph{i.e.,} $1760$ subjects) is used for estimating the model parameters and 33\% of the sample  (\emph{i.e.,} $866$ subjects) is used for investigating the performances of the models. The smoothing is performed on the continuous variables with a Gaussian kernel and a bandwidth $h=n^{-1/5}$.


\paragraph{Results}
We present the main results of the application. Details used for the results interpretation are presented in Appendix~\ref{app:appli}. 
We consider proposed approach in a semi-parametric framework with a quadratic loss.  According to the evolution of the smoothed log-likelihood with respect to the number of classes (see Figure~\ref{fig:smoothed} in Appendix~\ref{app:appli}), the model  is considered with $K=3$ classes.

To investigate the relevance of the activity level for explaining high blood pressure, we consider three models with a quadratic loss: the proposed approach in a semi-parametric framework  (\emph{regquadUZ-K3}), regression model of $Y$  on $U$ (\emph{regquadU}) with a selection of variables according to AIC (two variables are removed by the criterion: \emph{alchohol} and \emph{smoke}), regression model of $Y$ on $(U^\top,X^\top)$  (\emph{regquadUX}) with a selection of variables according to AIC (six variables are selected by the criterion: \emph{gender}, \emph{age}, \emph{obesity}, \emph{sleep}, \emph{cholesterol} and the binary variable indicating whether the subject usually travels on foot or by bike). Considering the activity levels seems to be relevant for explaining high blood pressure, since the MSE of prediction obtained on the testing samples are 122.55, 122.72 and 122.81 for \emph{regquadUZ-K3}, \emph{regquadUX} and \emph{regquadU} respectively. Thus, the approach permits to summarize the information about the physical activity and slighly improve the prediction accuracy. Note that a Shapiro-Wilk's normality test done on the residuals of \emph{regquadUZ-K3} has a pvalue less than $10^{-4}$ for both of the learning and testing sample. Thus, the semi-parametric approach avoids the normality assumption which is not relevant for the residuals.

To prevent the variability due to outliers, we fit the proposed approach in a semi-parametric framework with the median loss and the logcosh loss. Again, evolution of the smoothed log-likelihood with respect to the number of classes leads us to consider $K=3$ classes for both losses. We now compare the results obtained by the proposed method with $K=3$ classes in a semi-parametric framework with a quadratic loss, median loss (\emph{regmedUZ-K3}) and logcosh loss (\emph{reglogchUZ-K3}). The three models provided similar partition since the adjusted rand index between all the couples of partition is more than 0.85. The regression parameters are presented in Table~\ref{tab:param} of Appendix~\ref{app:appli}. The signs of the coefficients are the same for the three losses. It appears that being a woman prevents the high blood pressure while age, alcohol consumption, overweight, lack of sleeping and cholesterol increase high blood pressure. One can be surprised that results claim that smoking limits the risk of high blood pressure, but this effect has already been brought out in \citet{omvik1996smoking,li2017association}. Note that the robust methods detect a more significant effect of the alcohol, smoking and physical activity to high blood pressure. Moreover, they slightly improve the prediction accuracy because the MSE obtained on the testing sample is 122.44 and 122.48 for the median and the logcosh losses respectively.

We now interpret the clustering results provided by the median loss. Class 1 ($\pi_1=0.15$) is the smallest class and contains the subjects having recreational physical activities, traveling by foot or by bike, having no physical activity at work and spending few hours by watching screens. Class 2 ($\pi_2=0.44$) groups the subjects having few physical activities. Class 3 ($\pi_3=0.44$) groups the subjects having high physical activities at work. These results show that having moderate physical activities (recreational activities, traveling by bike or foot, not spending many hours by watching screens)  prevents the high blood pressure.

\section{Conclusion} \label{sec:conclusion}
In this paper, we propose an alternative to the two-step approach that starts by summarizing some observed variables by  clustering and then fitting a prediction model using the estimator of the partition as a covariate. Our proposition consists of simultaneously performing the clustering and the estimation of the prediction model to improve the accuracy of the partition and of the regression parameters. This approach can be applied to a wide range of regression models. 
Our proposition can be applied in a parametric and semi-parametric framework. 
We advise using the semi-parametric approach to avoid bias in the estimation (due to bias in the distribution modeling). 

The quality of the prediction could be used as a tool for selecting  the number of components and   bandwidth, for   semi-parametric mixtures. Like in any regression problem, this criterion can also be used for selecting the variables (in the regression part but also in the clustering part). Thus, taking into account the regression is important in model selection for semi-parametric mixtures. Moreover, this could allow for a variable selection in clustering while this approach is only used in a parametric framework \citep{tadesse2005bayesian,raftery2006variable}. 

The semi-parametric approach has been presented by assuming that the components are products of univariate densities. However, the proposed approach can also be used by considering location scale symmetric distributions \citep{hunterAOS2007} or by  incorporating an independent component analysis  structure \citep{zhu2019clustering}. 
Moreover, we can easily relax the assumption that $(X^\top,Z^\top)$ is independent of $U$. The crucial assumption of the model is the conditional independence of $Y$ and $X$ given $(Z^\top,U^\top)$.

This approach has been introduced by considering only one latent categorical variable. However, more than one latent categorical variable explained by different sub-groups of variables of $X$ could be considered. This extension is straightforward if the different sub-groups of variables of $X$ are known. However,  the case of the sub-groups of variables are also estimated (see the case of multiple partitions in clustering; \citet{MARBAC2019167}) could be considered in future work. 


 \bibliographystyle{apalike}
\bibliography{biblio}

\appendix
\section{Technical details}
\subsection{Proofs of the Lemmas}
\begin{proof}[Proof of Lemma~\ref{lemma:eqmoment}]
From \eqref{eq:condmoment}, noting that $Y\perp X\mid Z$, we have for all $k=1,\ldots,K$
\begin{align*}
&\mathbb{E}[\rho(Y - U^\top\gamma - \delta_k)| U=u,Z_k=1]=0\\
\Leftrightarrow\quad & \forall x,\; \int \rho(Y - u^\top\gamma - \delta_k) p(y|Z_k=1,u,x)dy=0. \\
\end{align*}
Thus, noting that $Z \perp U \mid X$ and that $f_k(x)>0$ implies $\mathbb{P}(Z_k=1|x,u)>0$, we have, that for $k=1,\ldots,K$
\begin{align*}
&\mathbb{E}[\rho(Y - U^\top\gamma - \delta_k)| U=u,Z_k=1]=0\\
\Leftrightarrow\quad & \forall x,\; \mathbb{P}(Z_k=1|x,u) \int \rho(Y - u^\top\gamma - \delta_k) p(y|Z_k=1,u,x)dy=0 \\
\Leftrightarrow\quad & \forall x,\;  \int \rho(Y - u^\top\gamma - \delta_k) \mathbb{P}(Z_k=1|y,x,u) p(y|u,x)dy=0 
\end{align*}
Proof is completed by noting that $\mathbb{P}(Z_k=1|y,x,u) =\mathbb{P}(Z_k=1|y,x)=r_k^{X,Y}(x,y)$ and that model identifiability involves that $\mathbb{E}[\rho(Y - U^\top\gamma - \delta_k)| U=u,Z_k=1]=0$ only for the true parameter $\beta$.
\end{proof}
 
\begin{proof}[Proof of Lemma~\ref{lemm:resasymptotic}]
Statement 1 holds because $Y$ is discriminative (\emph{i.e.,} its conditional distribution given $Z_k$ is different to its unconditional distribution). Thus, the optimal classification error based on $X$ is strictly more than the classification error based on $(X^\top,Y)$. 
Statement 2 is obtained by considering the quadratic loss and noting that $\mathbb{E}[Y|U=u,X=x,Z_k=1]=u^\top\gamma + \delta_k$. Thus, we have $(\tilde\gamma,\tilde\delta)^\top$ that statistifies
$$
\left\{ 
\begin{array}{rl}
\sum_{\ell=1}^K \Delta_{k\ell} \int g(u) (u^\top(\gamma - \tilde \gamma) + \delta_\ell - \tilde\delta_k)du & \forall k=1,\ldots,K\\
\sum_{k=1}^K \sum_{\ell=1}^K \Delta_{k\ell} \int g(u) u (u^\top(\gamma - \tilde\gamma) + \delta_\ell -\tilde\delta_k)du & 
\end{array}
\right.
$$
where $g(u)$ is the density of $U$. Thus, we have
$$
\left\{ 
\begin{array}{rl}
\tilde\delta_k = \mathbb{E}[U]^\top (\gamma - \tilde\gamma) + \frac{\sum_{\ell=1}^K \Delta_{k\ell} \delta_\ell}{\sum_{v=1}^K \Delta_{kv}} & \forall k=1,\ldots,K\\
\text{Var}(U) ( \gamma- \tilde \gamma) + \mathbb{E}[U] \sum_{k=1}^K \sum_{\ell=1}^K \Delta_{k\ell}(\delta_\ell -\tilde\delta_k)= 0.
\end{array}
\right.
$$
Substituting, in the second line, $\tilde\delta_k $ by its definition, and noting that $\text{Var}(U)$ has only non-zero eigen values, we obtain
$$
\left\{ 
\begin{array}{rl}
\tilde\delta_k =   \frac{\sum_{k=1}^K \Delta_{k\ell} \delta_\ell}{\sum_{v=1}^K \Delta_{kv}} & \forall k=1,\ldots,K\\
\tilde \gamma=\gamma.
\end{array}
\right.
$$
\end{proof}
 \subsection{EM algorithm for MLE estimation of the parametric simultaneous approach}\label{sec:EM}
 Starting from the initial point $\theta^{[0]}$, this iterative algorithm alternates between the E-step and M-step. Its iteration $[r]$ is defined by
\begin{itemize}
\item \textbf{E-step} Computation of the expectation of the complete-data log-likelihood
$$t_{ik}(\theta^{[r-1]})=\frac{\pi_k^{[r-1]} f_k(x_i;\alpha_k^{[r-1]}) f_\varepsilon(y_i -  u_i^\top \gamma^{[r-1]} - \delta_k^{[r-1]})}{\sum_{\ell=1}^K \pi_\ell^{[r-1]} f_\ell(x_i;\alpha_\ell^{[r-1]}) f_\varepsilon(y_i -  u_i^\top \gamma^{[r-1]} - \delta_\ell^{[r-1]})}.$$
\item \textbf{M-step} Maximization of the expectation of the complete-data log-likelihood
$$\pi_k^{[r]} = \frac{1}{n}\sum_{i=1}^n t_{ik}(\theta^{[r-1]}),\;\alpha_k^{[r]} = \argmax_{\alpha_k} \sum_{i=1}^n t_{ik}(\theta^{[r-1]}) \ln f_k(x_i;\alpha_k),$$
$$\beta^{[r]} =\argmax_{\beta} \sum_{i=1}^n \sum_{k=1}^K  t_{ik}(\theta^{[r-1]}) \ln f_\varepsilon(y_i -  u_i^\top \gamma - \delta_k).$$
\end{itemize}
The two maximizations at the M-step are standard. Indeed, the maximization problem provided by the computation of $\alpha_k^{[r]}$ already appears when the clustering is only made on $\rX$. Moreover, the maximization problem provided by the computation of $\beta^{[r]}$ is the solution of the regression problem with weighted observations.
 
\section{Details about the high blood pressure prevention data set}\label{app:appli}

\begin{figure}[htp]
\begin{center}
\includegraphics[scale=0.3]{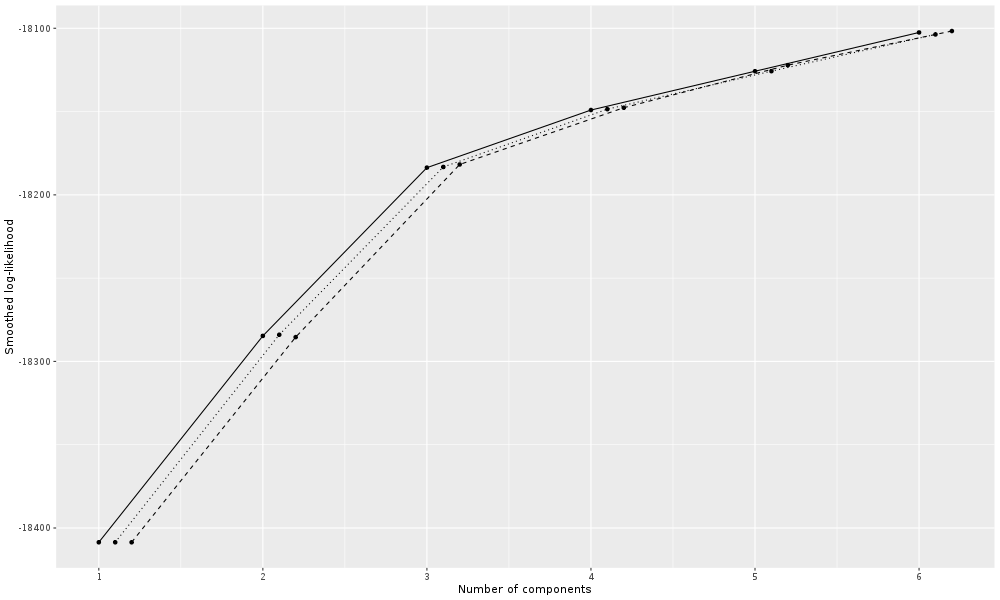}
\caption{Value of the smoothed log-likelihood function obtained for different number of classes by the quadratic loss (in plain line), the median loss (in dotted line) and the logcosh loss (in dashed line). Note that the x-axis has been drift for the median and logcosh losses for ease of reading.}\label{fig:smoothed}
\end{center}
\end{figure}

\begin{table}[htp]
\begin{center}
\begin{tabular}{cccc}
\hline
& \emph{regquadUZ-K3} & \emph{regmedUZ-K3} & \emph{reglogchUZ-K3} \\
\hline
Gender (woman)   &  -3.07 & -3.22 & -3.13\\
Age  &    0.24 &  0.23&   0.23\\
Alcohol (no)   &  -1.47 & -2.48 & -2.48\\
Obesity (yes) &     3.00  & 2.72 &  2.72\\
Sleep  &   -0.48 & -0.39 & -0.31\\
Smoke (no) &     0.09  & 0.42  & 0.51\\
Cholesterol &     0.03  & 0.03 &  0.03\\
Class 1 & 60.28 & 59.04 & 58.58\\
Class 2 & 60.52 & 59.75 & 59.48\\
Class 3 & 61.09 & 60.14 & 59.62\\
\hline
\end{tabular}
\end{center}
\caption{Estimators of the regression parameters given by the proposed semi-parametric approach with three losses.} \label{tab:param}
\end{table}

\begin{table}[ht]
\begin{center}
\begin{tabular}{rrr}
  \hline
 & yes & no \\ 
  \hline
class 1 & 0.03 & 0.97 \\ 
  class 2 & 0.10 & 0.90 \\ 
  class 3 & 0.99 & 0.01 \\ 
   \hline
\end{tabular}
\end{center}
\caption{Estimated probabilities per classes for the levels of the variable PAQ605 (vigorous work activity)} 
\end{table}
\begin{table}[ht]
\centering
\begin{tabular}{rrr}
  \hline
 & yes & no \\ 
  \hline
class 1 & 0.29 & 0.71 \\ 
  class 2 & 0.23 & 0.77 \\ 
  class 3 & 0.94 & 0.06 \\ 
   \hline
\end{tabular}
\caption{Estimated probabilities per classes for the levels of the variable PAQ620 (moderate work activity)} 
\end{table}
\begin{table}[ht]
\centering
\begin{tabular}{rrr}
  \hline
 & yes & no \\ 
  \hline
class 1 & 0.45 & 0.55 \\ 
  class 2 & 0.29 & 0.71 \\ 
  class 3 & 0.35 & 0.65 \\ 
   \hline
\end{tabular}
\caption{Estimated probabilities per classes for the levels of the variable PAQ635 (usal travels on foot or by bike)} 
\end{table}
\begin{table}[ht]
\centering
\begin{tabular}{rrr}
  \hline
 & yes & no \\ 
  \hline
class 1 & 0.57 & 0.43 \\ 
  class 2 & 0.07 & 0.93 \\ 
  class 3 & 0.42 & 0.58 \\ 
   \hline
\end{tabular}
\caption{Estimated probabilities per classes for the levels of the variable PAQ650 (vigorous recreational activities)} 
\end{table}
\begin{table}[ht]
\centering
\begin{tabular}{rrr}
  \hline
 & yes & no \\ 
  \hline
class 1 & 0.67 & 0.33 \\ 
  class 2 & 0.23 & 0.77 \\ 
  class 3 & 0.52 & 0.48 \\ 
   \hline
\end{tabular}
\caption{Estimated probabilities per classes for the levels of the variable PAQ665 (moderate recreational activities)} 
\end{table}
\begin{table}[ht]
\centering
\begin{tabular}{rrrrrrrr}
  \hline
 & never & less than 1h & 1h & 2h & 3h & 4h & 5h and more \\ 
  \hline
class 1 & 0.02 & 0.15 & 0.24 & 0.25 & 0.18 & 0.08 & 0.07 \\ 
  class 2 & 0.02 & 0.10 & 0.13 & 0.26 & 0.17 & 0.12 & 0.20 \\ 
  class 3 & 0.01 & 0.14 & 0.22 & 0.33 & 0.13 & 0.10 & 0.07 \\ 
   \hline
\end{tabular}
\caption{Estimated probabilities per classes for the levels of the variable PAQ710 (time spent watching TV)} 
\end{table}
\begin{table}[ht]
\centering
\begin{tabular}{rrrrrrrr}
  \hline
 & never & less than 1h & 1h & 2h & 3h & 4h & 5h and more \\ 
  \hline
  class 1 & 0.08 & 0.22 & 0.27 & 0.20 & 0.10 & 0.05 & 0.08 \\ 
  class 2 & 0.30 & 0.24 & 0.13 & 0.12 & 0.07 & 0.05 & 0.09 \\ 
  class 3 & 0.21 & 0.31 & 0.21 & 0.14 & 0.04 & 0.03 & 0.06 \\ 
   \hline
\end{tabular}
\caption{Estimated probabilities per classes for the levels of the variable  PAQ715 (time spent using a computer)} 
\end{table}
\end{document}